\documentclass[fleqn,usenatbib,useAMS]{mnras}
\usepackage{mathtools}
\usepackage{txfonts}

\usepackage[T1]{fontenc}
\usepackage{ae,aecompl}

\usepackage{color}
\usepackage{graphicx}

\usepackage{amsmath,amssymb,amsfonts}

\def\dobs{d_{\rm obs}}
\def\apj{Astrophys. J.}

\def\apjl{Astrophys. J. Lett.}
\def\aap{Astronomy and Astrophysics}

\def\mnras{Mon. Not. R. Astron. Soc.}

\newcommand{\oldtext}[1]{}

\title[Tidal encounters and the dimming of Betelgeuse]{Did a close tidal encounter cause the Great Dimming of Betelgeuse?}

\author[Aronson, Baumgarte \& Shapiro]{Hailey Aronson,$^{1}$ Thomas W.\ Baumgarte,$^{1}$ and Stuart L.\ Shapiro$^{2,3}$ \\ \\
$^{1}$Department of Physics and Astronomy, Bowdoin College,
 Brunswick, ME 04011 \\
$^{2}$Department of Physics, University of Illinois at Urbana-Champaign, Urbana, IL 61801\\
$^{3}$Department of Astronomy and NCSA, University of Illinois at Urbana-Champaign, Urbana, IL 61801}

\date{Accepted XXX. Received YYY; in original form ZZZ}

\pubyear{2022}

\begin{document}
\label{firstpage}
\pagerange{\pageref{firstpage}--\pageref{lastpage}}
\maketitle

\begin{abstract}
We assess whether gravity darkening, induced by a tidal interaction during a stellar fly-by, might be sufficient to explain the Great Dimming of Betelgeuse. Adopting several simple approximations, we calculate the tidal deformation and associated gravity darkening in a close tidal encounter, as well as the reduction in the radiation flux as seen by a distant observer.  We show that, in principle, the duration and degree of the resulting stellar dimming can be used to estimate the minimum pericenter separation and mass of a fly-by object, which, even if it remains undetected otherwise, might be a black hole, neutron star, or white dwarf.  Our estimates show that, while such fly-by events may occur in other astrophysical scenarios, where our analysis should be applicable, they likely are not large enough to explain the Great Dimming of Betelgeuse by themselves.
\end{abstract}

\begin{keywords}
stars: peculiar (except chemically peculiar) -- stars: variables: general -- transients: tidal disruption events
\end{keywords}

%==================================================
\section{Introduction}
\label{sec:intro}
%==================================================

The ``Great Dimming" of Betelgeuse, observed from late 2019 into early 2020, reduced its visual brightness by about one apparent magnitude, corresponding to a reduction in intensity by about 60\%. A number of different effects have been suggested that might have caused this dimming, including changes in the star's photosphere \citep{DhaMSBMMWZ20}, a critical transition in the pulsation dynamics \citep{GeoKMA20}, strong outflows, resulting from a conjunction of shock waves and convective motion and leading to an increase in molecular opacity \citep{Kraetal21}, the appearance of a large dark spot \citep{AleZDLLH21}, as well as obscuring by a dust cloud expelled during a recent mass-loss episode \citep{LevM20,Dupetal20,Monetal21}.

\begin{figure}
    \centering
    \includegraphics[width = .45 \textwidth]{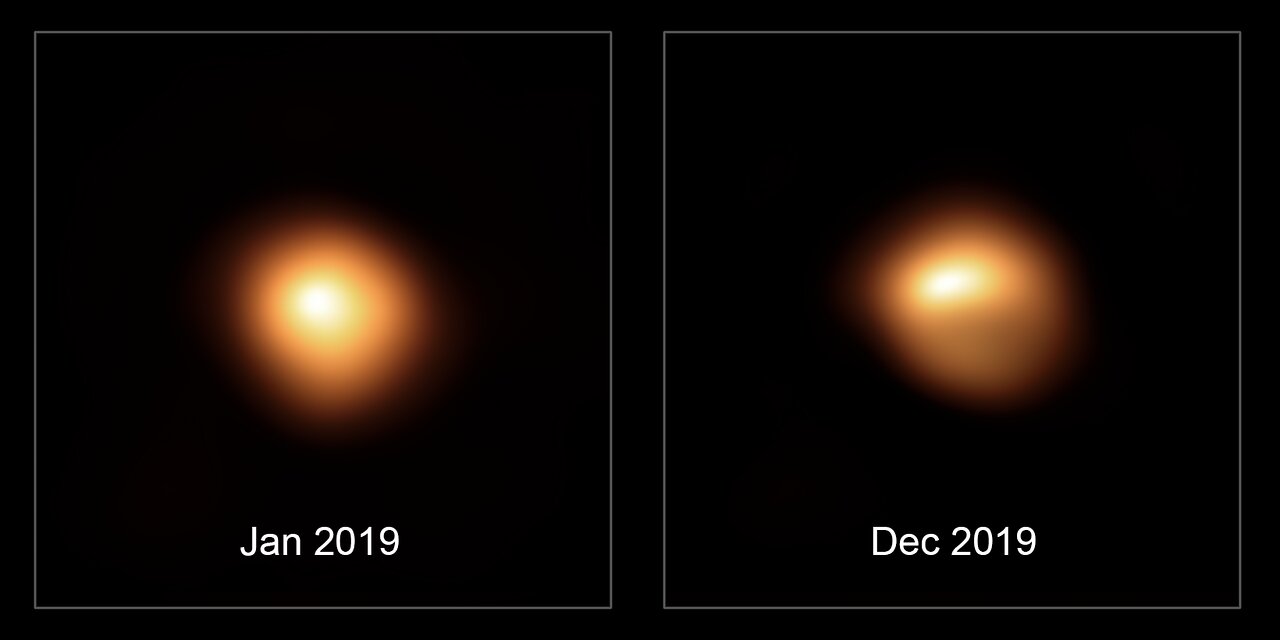}
    \caption{Images of Betelgeuse before (left panel) and during (right panel) the Great Dimming event.  Note that, during the dimming, Betelgeuse seems to be deformed, with its elongated side appearing dimmer than the rest of the star.  Credit: ESO/M.~Montarg\`es {\it et.al.} \citep[see][]{MonCKF20}.}
    \label{fig:eso_image}
\end{figure}

Resolved images of Betelgeuse also show that the star seemed deformed during this dimming event, with the elongated side appearing dimmer than the rest of the star (\cite{MonCKF20}; see Fig.~\ref{fig:eso_image}).  While this appearance is consistent with some of the effects listed above, this image could easily be interpreted as showing a tidal deformation.  An unseen object -- possibly a black hole, neutron star or white dwarf -- close to Betelgeuse's surface could raise a tidal bulge, which, due to gravity darkening, would appear less bright, and hence cause Betelgeuse's dimming.

Gravity darkening, first studied by \citet{Zei24a,Zei24b,Zei24c}, results from the emerging radiative flux being related to the effective gravitational force on the stellar surface (see, e.g., \cite{Zei24a,Cha33a}, as well as \cite{Tas00} and \cite{KipWW12} for textbook treatments).  Two different effects may reduce the effective gravitational potential and may hence lead to a dimming of the star, namely rotation (e.g.~\cite{Zei24a,Zei24b,Cha33a,CraO95,BauS99}) and tidal interaction with a companion (e.g.~\cite{Zei24c,Cha33b,Cha33c,WhiBS12}).  
Effects of gravity darkening have been observed in resolved, interferometric images of rotating stars (e.g.~\cite{Monetal07,Zhaetal09}, and they have been used to model light-curves from close binary systems (e.g.~\cite{RafT80,DjuRRGEP03,DjuRRGEP06}).  Based on these observations several authors have also suggested improvements of the simple von Zeipel relations (see, e.g.~\cite{EspR11,Cla16,ZorREDR17} and references therein).

Building on the above treatments we explore in this short paper whether the effects of such gravity darkening, induced by the tidal interaction with a ``fly-by" object in hyperbolic orbit, might be large enough to explain the Great Dimming of Betelgeuse.  We adopt a number of crude approximations to model the tidal deformation of a star in response to an object passing by the star, to compute the flux emitted from the star's deformed surface, and to determine the intensity as seen by an observer at large distance.  We show that the duration and degree of the darkening event may provide estimates for the minimum pericenter separation and mass of the fly-by object.  While these effects may well be observable in other events, we find that, at least according to our leading-order estimates, they likely are not sufficiently large to explain the Great Dimming of Betelgeuse.  

We finally note that, even though a close tidal encounter does not appear viable to explain the Great Dimming by means of gravity darkening alone, it remains possible, of course, that such an interaction may either have caused or contributed to the launching of a stellar outburst or wind, which has been invoked to explain the obscuring of Betelgeuse by a dust cloud \citep{LevM20,Dupetal20,Monetal21}.

%==================================================
\section{Timescales}
\label{sec:timescales}
%==================================================

We assume that an object, which we will refer to as the companion, is in a hyperbolic orbit about the primary star, Betelgeuse. We will refer to the smallest separation between the two objects, i.e., the pericenter separation, as $s$.  Neglecting the kinetic energy when the two objects are at large separation, the relative speed at pericenter is given by
\begin{equation} \label{v_s}
    \varv_s \simeq \left( \frac{2 G M}{s} \right)^{1/2} = \left( \frac{2 G M_p}{s} \right)^{1/2} (1 + q)^{1/2},
\end{equation}
where $M = M_p + M_c = M_p ( 1 + q)$ is the total mass, with $M_p$ and $M_c$ the masses of the primary and companion, respectively, and where we have defined the mass ratio
\begin{equation}
    q \equiv \frac{M_c}{M_p}.
\end{equation}
The duration of this fly-by event is then approximately
\begin{equation} \label{fb}
    \tau_{\rm fb} \simeq \frac{s}{\varv_s} \simeq \left( \frac{s^3}{2 G M_p} \right)^{1/2} (1 + q)^{-1/2}.
\end{equation}
The timescale for the stellar atmosphere to respond to this perturbation can be estimated from the dynamical timescale 
\begin{equation}\label{dyn}
    \tau_{\rm dyn} \simeq \left( \frac{R_0^3}{G M_p} \right)^{1/2},
\end{equation}
where $R_0$ is the star's unperturbed radius.  We can also combine Eqs.~(\ref{fb}) and (\ref{dyn}) to obtain
\begin{equation} \label{pericenter}
\frac{\tau_{\rm fb}}{\tau_{\rm dyn}} = 2^{-1/2} (1 + q)^{-1/2} \sigma^{3/2},
\end{equation}
which shows that $\tau_{\rm fb}$, in units of the star's dynamical timescale $\tau_{\rm dyn}$, provides one relation for the mass ratio $q$ and the dimensionless pericenter separation
\begin{equation} \label{sigma}
    \sigma \equiv \frac{s}{R_0}.
\end{equation}
We will return to the relation (\ref{pericenter}) in Section \ref{sec:betelgeuse} below, after finding a second relation for $q$ and $\sigma$ in Section \ref{sec:observer} (see Eq.~\ref{F_rel} below).

%==================================================
\section{Hydrostatic equilibrium}
\label{sec:hydrostatic}
%==================================================

In the following we will treat the tidal deformation caused by the companion quasi-statically, i.e.~ignoring all dynamical effects.  This approximation becomes increasingly accurate as $\tau_{\rm fb} \gg \tau_{\rm dyn}$, or, from (\ref{pericenter}), as $s \gg R_0$.

\begin{figure}
    \centering
    \includegraphics[width = 0.48 \textwidth]{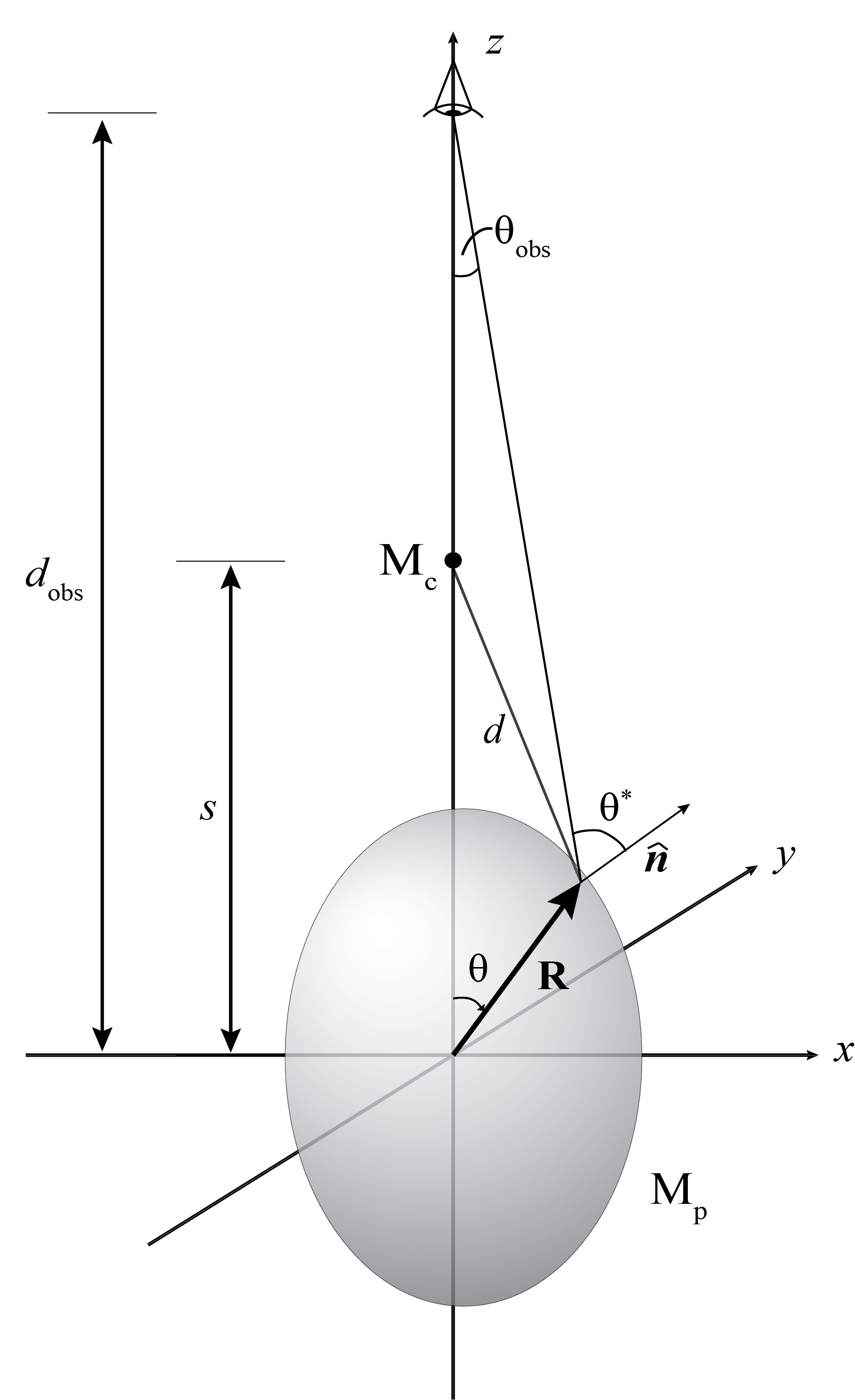}
    \caption{A sketch of the tidally distorted primary $M_p$ during a fly-by encounter with a companion $M_c$, with an observer at large distance $d_{\rm obs}$.  Our leading-order analysis applies equally for a fly-by object on the negative $z$-axis, i.e.~in opposition rather than conjunction.}
    \label{fig:sketch}
\end{figure}

In hydrostatic equilibrium we have
\begin{equation} \label{hydro1}
    \frac{1}{\rho} \, \nabla P = - \nabla \Phi,
\end{equation}
where $P$ is the pressure, $\rho$ the density, and $\Phi = \Phi_p + \Phi_c$ the Newtonian gravitational potential. Here we neglect effects of rotation, as is appropriate for Betelgeuse. We now adopt the Roche model for the outer atmosphere of the primary, whereby we assume that most of the star's mass resides in its unperturbed interior. We may then approximate the gravitational potential in the atmosphere by that of a point mass,
\begin{equation} \label{pot_primary}
    \Phi_p \simeq - \frac{GM_p}{r},
\end{equation}
where $r$ is the distance from the center of the primary. Modeling the companion as a point mass as well, its potential is $\Phi_c =- M_c/d$, where $d = (r^2 + s^2 - 2 r s \cos \theta)^{1/2}$ is the distance to the companion (see Fig~\ref{fig:sketch}). We now expand 
\begin{equation} \label{expansion}
    \frac{1}{d} = \frac{1}{s} \sum_{\ell=0}^\infty \left( \frac{r}{s} \right)^\ell P_\ell(\cos \theta),
\end{equation}
where the $P_\ell(\cos \theta)$ are the Legendre polynomials of order $\ell$.  The monopole term $\ell = 0$ in (\ref{expansion}) is a constant and can therefore be ignored.  The dipole term $\ell = 1$ represents a constant force exerted by the companion on the primary, causing a slight shift in its center of mass; we will ignore this term as well since we only are interested in the primary's tidal deformation. Thus, this deformation is dominated by the quadrupole term, which we therefore employ for the companion's gravitational potential,\footnote{It would be straightforward to include the higher-order terms $\ell > 2$ as well.  However, these become important only when $r \simeq s$, in which case our assumption of hydrostatic equilibrium is likely to fail anyway.}
\begin{equation} \label{phi_c}
    \Phi_c \simeq - \frac{G M_c r^2}{s^3} \, P_2(\cos \theta) = - \frac{G M_c r^2}{2 s^3} \, \left( 3 \cos^2 \theta - 1 \right).
\end{equation}

Integrating (\ref{hydro1}) we obtain
\begin{equation} \label{first_int1}
    h + \Phi_p + \Phi_c = H
\end{equation}
where $h = \int dP / \rho$ is the specific enthalpy, and where $H$ is a constant of integration.  Evaluating (\ref{first_int1}) for the {\it unperturbed} star (i.e.~$\Phi_c = 0$) at the primary's surface, where $h = 0$,  we have $H = - G M_p /R_0$. To determine $H$ in the presence of the companion we evaluate (\ref{first_int1}) in the primary's center, which we assume is largely unperturbed, and where $\Phi_c = 0$. The constant $H$ thus retains its unperturbed value, $H = -  G M_p / R_0$, even in the presence of a companion (see also \cite{WhiBS12} for discussion and references).

We may now find the stellar surface $R = R(\theta)$ by setting $h=0$ in (\ref{first_int1}), or equivalently, by solving
\begin{align} \label{first_int2}
S(\zeta, \theta) & \equiv \frac{R_0}{G M_p} \left( \frac{G M_p}{R_0} + \Phi_p + \Phi_c \right) \nonumber  \\
& =  1 - \frac{1}{\zeta} - \frac{q \zeta^2}{2 \sigma^3} \left(3 \cos^2 \theta - 1 \right) = 0
\end{align}
where we have defined the dimensionless radius
\begin{equation}
    \zeta \equiv \frac{R}{R_0}.
\end{equation}
For given values of $q$, $\sigma$, and $\theta$ we can now find the location of the primary's surface by solving (\ref{first_int2}) for $\zeta = \zeta(\theta)$.

The limiting separation occurs when the the pressure $P$, at the point on the primary's surface facing the companion (i.e.~for $\theta = 0$), no longer increases inwards, indicating the onset of mass transfer and possible tidal disruption.  Setting the gradient of the potential $\Phi$ to zero for $\theta = 0$ yields $\sigma_{\rm lim} = (2 q)^{1/3} \, \zeta_{\rm lim}$, and inserting this expression into (\ref{first_int2}) we find $\zeta_{\rm lim} = 3/2$ and hence $\sigma_{\rm lim} =  (2 q)^{1/3} \, (3 / 2)$.  For $q < 1/2$, however, this results in $\sigma_{\rm lim} < \zeta_{\rm lim}$, suggesting that the companion has been absorbed by the primary, and clearly violating our assumptions.  This result is an artifact of us having neglected the higher-order terms in (\ref{expansion}), which otherwise would have lead to formation of a cusp prior to binary contact.  Therefore, for $q < 1/2$, we simply adopt $\sigma_{\rm lim} = \zeta_{\rm lim}$ (i.e. $s=R$ at $\theta =0$) which results in $\sigma_{\rm lim} = 1 + q$, and we accordingly truncate our calculations at\footnote{A more realistic treatment would invoke the onset of Roche-lobe overflow as the limit of these sequences, see, e.g., \cite{Egg83}.}
\begin{equation} \label{sigma_lim}
    \sigma_{\rm lim} = \left\{ \begin{array}{ll}
         \displaystyle (3/2) \, (2q)^{1/3}~~~  & q \geq 1/2  \\
         1 + q & q < 1/2.
    \end{array}
    \right.
\end{equation}
We caution, however, that the effects of having dropped the higher-order terms in (\ref{expansion}), as well as those of treating the problem statically rather than dynamically, become increasingly large as $\sigma \rightarrow \zeta$.

%==================================================
\section{Emergent flux and intensity}
\label{sec:flux}
%==================================================

In the diffusion approximation in the radiating surface layer we may write the flux ${\bf F}$ as 
\begin{equation}
    {\bf F} = - \frac{c}{3 \kappa \rho} \nabla U,
\end{equation}
where $\kappa$ is the Rosseland mean opacity, $U = 3 P_{\rm rad}$ the energy density of the radiation, and $P_{\rm rad}$ the radiation pressure.  We now write the total pressure as $P = P_{\rm gas} + P_{\rm rad}$ where $P_{\rm gas}$ is the gas pressure, and define $\beta \equiv P_{\rm gas} / P_{\rm rad}$.  Supergiants like Betelgeuse are dominated by radiation pressure, i.e.~$\beta \ll 1$.  We therefore neglect effects of changes in the small quantity $\beta$ and obtain 
\begin{equation}
    {\bf F} = - \frac{c}{\kappa \rho} \nabla P_{\rm rad}  \simeq - \frac{c}{(1 + \beta) \kappa \rho} \, \nabla P = \frac{c}{(1 + \beta) \kappa} \, \nabla (\Phi_p + \Phi_c),
\end{equation}
where we have used (\ref{hydro1}) in the last equality.  Evaluating the gradient of the gravitational potential $\Phi = \Phi_p + \Phi_c$ we have
\begin{equation} \label{flux}
    {\bf F} = \frac{c}{(1 + \beta) \kappa} \frac{G M_p}{R_0^2} \left\{
    f^{\hat r} \, \hat {\bf r} + f^{\hat \theta} \, \hat {\bf \theta} \right\} 
\end{equation}
where 
\begin{align}
%    f^{\hat r} & \equiv \frac{1}{\zeta^2} + \frac{q}{\delta^3} \, \left(\zeta - \sigma \cos \theta \right) \\
%    f^{\hat \theta} & \equiv \frac{q}{\delta^3} \sigma \sin \theta 
        f^{\hat r} & \equiv \frac{1}{\zeta^2} - \frac{q \zeta}{\sigma^3} \, \left(3 \cos^2 \theta - 1 \right) \\
    f^{\hat \theta} & \equiv 3 \frac{q \zeta}{\sigma^3} \cos \theta \sin \theta 
\end{align}
are the components of the non-dimensional flux 
\begin{equation} \label{f_define}
    {\bf f} \equiv \frac{(1 + \beta) \, \kappa}{c} \, \frac{R_0^2}{G M_p} \,  {\bf F}.
\end{equation}
In the absence of a companion we have $q = 0$ and hence $\zeta = 1$.  In this case the flux reduces to 
\begin{equation} \label{F_emit_0}
    {\bf F}_0 = \frac{c}{(1 + \beta) \kappa} \frac{G M_p}{R_0^2}\, \hat {\bf r}
    = \frac{L_0}{4 \pi R_0^2} \, \hat {\bf r}, 
    \hfill \mbox{(unperturbed)}
\end{equation}
where we have introduced the unperturbed stellar luminosity
\begin{equation}
    L_0 \equiv \frac{4 \pi G c M_p}{(1 + \beta)\kappa} 
\end{equation}
in the last equality.  Note that we may therefore rewrite the non-dimensional flux (\ref{f_define}) as ${\bf f} = {\bf F} / F_0$.

We next compute the (energy-integrated) intensity $I$ from the relation 
\begin{equation} \label{F_ito_I}
    F(\theta) = \int_{\rm outward} I(\theta,\theta^*) \cos \theta^* d \Omega^*,
\end{equation}
where $F(\theta)$ is the magnitude of the flux (\ref{flux}), and where we are allowing the intensity to depend on the direction of emission $\theta^*$, measured with respect to the normal $\hat {\bf n}$ to the stellar surface  as shown in Fig.~\ref{fig:sketch}, in addition to the position on the surface, i.e.~$\theta$.  We note that the flux ${\bf F}$ is aligned with $\hat {\bf n}$.

%==================================================
%\subsection{No limb darkening}
%\label{sec:nolimbdark}
%==================================================

In the {\em absence of limb darkening} we assume the radiation to be emitted isotropically in the outward direction, so that the intensity is independent of $\theta^*$.  Carrying out the integration in (\ref{F_ito_I}) then yields
\begin{equation} \label{I_no_ld}
    I^{\rm iso}(\theta) = \frac{F(\theta)}{\pi} = F_0 \frac{f(\theta)}{\pi} 
    \hfill \mbox{(no limb darkening),} 
\end{equation}
where we have used $F(\theta) = F_0 f(\theta)$ in the last equality, and where $f(\theta)$ is the magnitude of the vector ${\bf f}$ defined in (\ref{f_define}).

Adopting the Eddington approximation for {\em limb darkening}, however, we assume that the the emitted intensity depends on the angle $\theta^*$ according to 
\begin{equation} \label{I_ld}
    I^{\rm ld}(\theta,\theta^*) = \frac{3}{5} \, \left( \cos \theta^* + \frac{2}{3} \right) I_0^{\rm ld}(\theta). \hfill \mbox{(limb darkening)}
\end{equation}
Inserting this expression into (\ref{F_ito_I}) shows that
\begin{equation}
    I_0^{\rm ld}(\theta) = \frac{5 F(\theta)}{4 \pi} = F_0 \frac{5 f(\theta)}{4 \pi}. 
    \hfill \mbox{(limb darkening)}
\end{equation}

%==================================================
\section{Observed brightness}
\label{sec:observer}
%==================================================

In order to estimate the largest possible reduction in brightness we now consider an observer at large distance in the $z$-direction.  The flux as measured by this observer can again computed from the intensity,
\begin{equation} \label{F_obs1}
    F_{\rm obs} = \int_{\rm outward} I \cos \theta_{\rm obs} d \Omega_{\rm obs}
    \simeq 2 \pi \int_{\rm outward} I(\theta, \theta^*) \, \theta_{\rm obs} d \theta_{\rm obs},
\end{equation}
where we have used $d \Omega_{\rm obs} = \sin \theta_{\rm obs} d \theta_{\rm obs} d \phi_{\rm obs}$, have carried out the integration over $\phi_{\rm obs}$, and have used $\theta_{\rm obs} \ll 1$ to approximate $\cos \theta_{\rm obs} \simeq 1$ and $\sin \theta_{\rm obs} \simeq \theta_{\rm obs}$.  We next express the angles $\theta_{\rm obs}$ and $\theta^*$ in terms of $\theta$.

From Fig.~\ref{fig:sketch} we first note that \begin{equation}
    \theta_{\rm obs} \simeq \frac{R \sin \theta}{\dobs} % = \frac{R_0 \zeta \sin \theta}{\dobs}
\end{equation}
and hence
\begin{equation} \label{dtheta_obs}
    d \theta_{\rm obs} \simeq \frac{1}{\dobs} \, \left( \sin \theta \frac{dR}{d\theta} + R \cos \theta \right) \, d \theta.
\end{equation}
Since, along the stellar surface we have $S(\zeta, \theta) = 0$ and hence
\begin{equation}
    dS = \frac{\partial S}{\partial \zeta} d \zeta + \frac{\partial S}{\partial \theta} d \theta = 0,
\end{equation} 
we may evaluate the term $dR/d\theta$ in (\ref{dtheta_obs}) as
\begin{equation}
\frac{d R}{d \theta} = 
R_0 \frac{d \zeta}{d \theta} = - R_0 \frac{\partial S / \partial \theta}{\partial S / \partial \zeta} = - R_0 \frac{\partial \Phi / \partial \theta}{\partial \Phi / \partial \zeta} = - R \frac{F^{\hat \theta}}{F^{\hat r}} =  - R_0 \zeta \frac{f^{\hat \theta}}{f^{\hat r}}
\end{equation}
Inserting the above into (\ref{F_obs1}) we obtain
\begin{equation}
    F_{\rm obs} = \frac{2 \pi R_0^2}{\dobs^2} \int_{\rm outward} I(\theta,\theta^*) \, \zeta^2 \sin \theta \left( \cos \theta - \frac{f^{\hat \theta}}{f^{\hat r}} \sin \theta  \right) d \theta.
\end{equation}

The angle $\theta^*$ corresponding to emission in the direction of the observer at a large distance on the (positive) $z$-axis can be found from 
\begin{equation}
    \cos \theta^* \simeq \hat {\bf n} \cdot \hat {\bf z} = \frac{1}{F} \, {\bf F} \cdot \hat {\bf z} = \frac{F^z}{F} = \frac{f^z}{f} = \frac{f^{\hat r} \cos \theta - f^{\hat \theta} \sin \theta}{f},
\end{equation}
where $\hat {\bf n}$ is the normal on the surface, which is aligned with the flux ${\bf F}$.  

In the absence of a companion we have already seen that the magnitude of the emitted flux is independent of the angle $\theta$.  In this case we also have $\theta^* = \theta$, $\zeta = 1 = f^{\hat r}$ and $f^{\hat \theta} =0$ and hence, not unexpectedly,
\begin{equation} \label{F_obs_0}
    F_{\rm obs,0} = \frac{2 \pi R_0^2}{\dobs^2} \int_0^{\pi/2} I(\theta, \theta) \sin \theta \cos \theta d \theta = \frac{R_0^2}{\dobs^2} \, F_0 \hfill \mbox{(unperturbed)}
\end{equation}
where we have used (\ref{F_ito_I}) in the second equality.  Note that the unperturbed result (\ref{F_obs_0}) (the simple inverse-square law for flux) holds independently of any assumptions on limb darkening.  

We now compute the {\em relative} observed flux from
\begin{equation} \label{F_rel}
    \frac{F_{\rm obs}}{F_{\rm obs,0}} = 2 \pi \int_0^{\pi/2} \frac{I(\theta,\theta^*)}{F_0} \, \zeta^2 \sin \theta \left( \cos \theta - \frac{f^{\hat \theta}}{f^{\hat r}} \sin \theta \right) d \theta,
\end{equation}
where the non-dimensional intensity $I(\theta,\theta^*)/F_0$ can be computed from (\ref{I_no_ld}) or (\ref{I_ld}) in terms of the magnitude $f$ of the non-dimensional flux ${\bf f}$ (see \ref{f_define}).  In (\ref{F_rel}), as well as in deriving (\ref{F_emit_0}), we also assumed that the opacity $\kappa$ and the pressure coefficient $\beta$ remain unaffected by the tidal deformation, so that they cancel out.  Given an observation of the relative flux, Eq.~(\ref{F_rel}) provides a second relation for $q$ and $\sigma$, which, at least in principle, can be used together with (\ref{pericenter}) to find $q$ and $\sigma$ individually.

\begin{figure}
    \centering
    \includegraphics[width = 0.48 \textwidth]{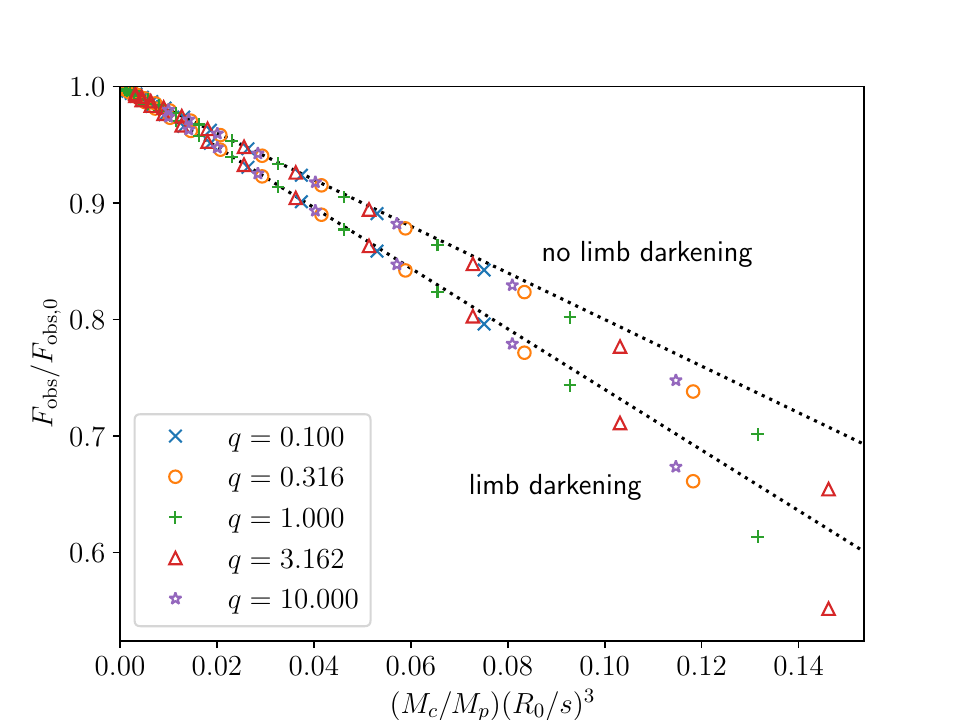}
    % Created with command make_plot() in betelgeuse_intensity_lo.py in Students/Hailey_Aronson
    \caption{Values of the relative observed flux (see Eq.~\ref{F_rel}) as seen by an observer at a large distance in the $z$-direction, as a function of the combination $q / \sigma^3 = M_c/M_p (R_0/s)^3$.  The dotted lines represent the linear results (\ref{F_rel_lin}).}
    \label{fig:intensity}
\end{figure}

For the unperturbed star, the relative observed flux always evaluates to unity, as expected.  By virtue of us having truncated the potential $\Phi_c$ after the quadrupole term in (\ref{phi_c}), all terms that deviate from the spherical, unperturbed values scale with the combination
\begin{equation}
    \frac{q}{\sigma^3} = \frac{M_c}{M_p} \left( \frac{R_0}{s} \right)^3.
\end{equation}
As a result, the relative observed brightness can be a function of the $q / \sigma^3$ only, rather than depending on $q$ and $\sigma$ individually.  This can be seen in Fig.~\ref{fig:intensity}, where we plot the relative observed flux, computed for different values of $q$ and $\sigma$, as a function of $q / \sigma^3$.  We include results both with and without limb darkening, and observe that the effects of gravity darkening are slightly larger when limb darkening is taken into account, as one might have expected.

It is also possible to expand the integrand in (\ref{F_rel}) in powers of $q / \sigma^3$, and integrate the individual terms separately, in which case we obtain
\begin{equation} \label{F_rel_lin}
     \frac{F_{\rm obs}}{F_{\rm obs,0}} = 1 - \alpha \frac{q}{\sigma^3} + {\mathcal O}\left((q/\sigma^3)^2 \right)
\end{equation}
with
\begin{equation}
    \alpha = \left\{
    \begin{array}{ll}
         2 & \mbox{no limb darkening} \\
         13/5 & \mbox{limb darkening}. 
    \end{array}
    \right.
\end{equation}
The dotted lines in Fig.~\ref{fig:intensity} show these linear results alone.

%==================================================
\section{Discussion and application to Betelgeuse}
\label{sec:betelgeuse}
%==================================================

\begin{figure*}
    \centering
    \includegraphics[width = 0.48 \textwidth]{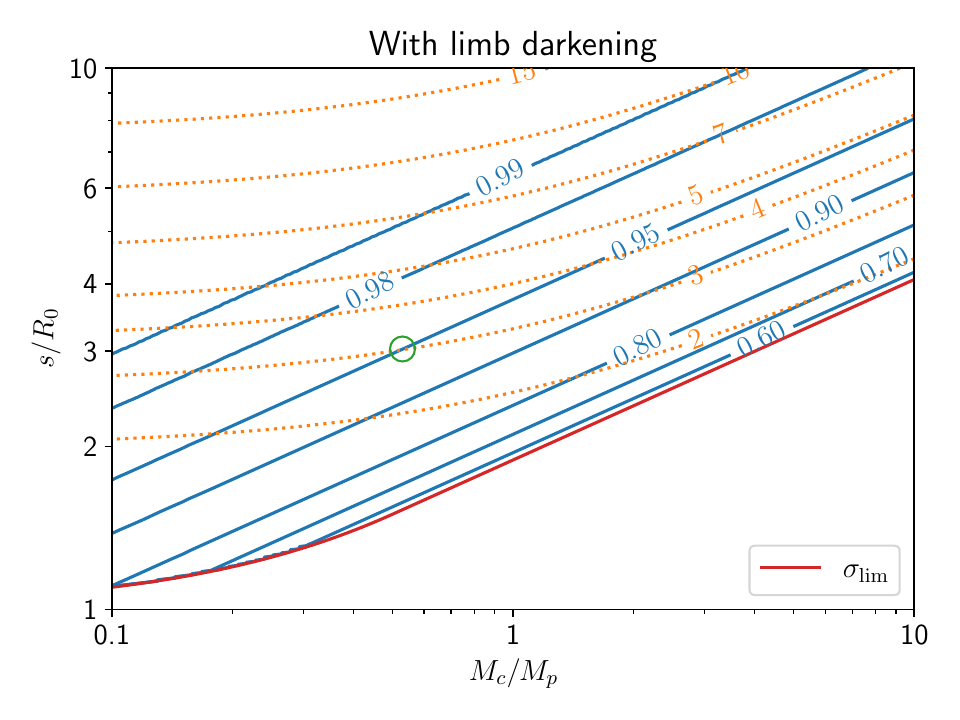}
    \includegraphics[width = 0.48 \textwidth]{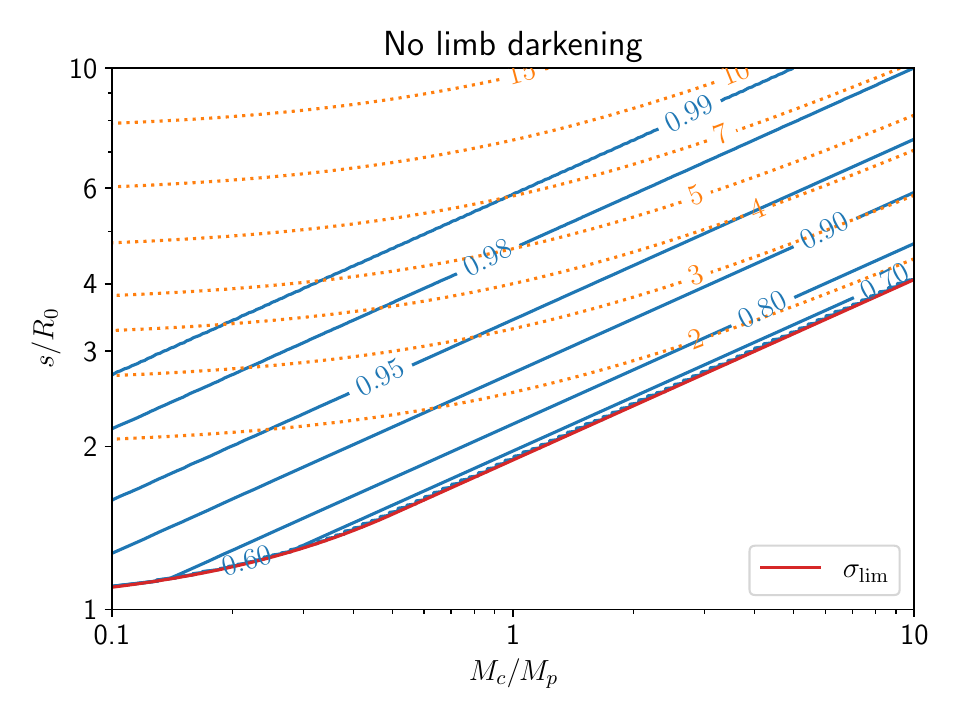}
    % Created with command contour_plot() in betelgeuse_intensity_lo.py in Students/Hailey_Aronson
    \caption{Contours of the relative observed flux (solid (blue) lines), see Eq.~(\ref{F_rel}) as seen by an observer at a large distance in the $z$-direction, as a function of mass ratio $q = M_c / M_p$ and relative companion separation $\sigma = s / R_0$.  The dotted (orange) lines show contours of the ratio between the fly-by timescale $\tau_{\rm fb}$ and the dynamical timescale $\tau_{\rm dyn}$, see Eq.~(\ref{pericenter}).  We show results with limb darkening in the left panel, and without limb darkening in the right panel.  The limiting companion separation $\sigma_{\rm lim}$ is given by Eq.~(\ref{sigma_lim}).  The open (green) circle in the left panel marks the specific example discussed in the text.}
    \label{fig:darkening}
\end{figure*}

In Fig.~\ref{fig:darkening} we show contours of the relative observed flux, Eq.~(\ref{F_rel}) as well as the ratio between the fly-by timescale $\tau_{\rm fb}$ and the dynamical timescale $\tau_{\rm dyn}$, Eq.~(\ref{pericenter}), both with and without limb darkening.  Since the former depends on $q / \sigma^3$ only, its contours appear as straight lines in the log-log plots of Fig.~\ref{fig:darkening}.  For large $q$, the latter also depends approximately on $q / \sigma^3$, so that both families of contours become nearly parallel in this limit.  This means that $q$ and $\sigma$ can no longer be determined from a measurement of $F_{\rm obs} / F_{\rm obs,0}$ and $\tau_{\rm fb} / \tau_{\rm dyn}$ in this limit.

For sufficiently small $q$, however, independent measurements of $F_{\rm obs} / F_{\rm obs,0}$ and $\tau_{\rm fb} / \tau_{\rm dyn}$ would provide estimates for the minimum mass ratio $q$ and pericenter separation $\sigma$.  As a concrete example, imagine a measurement of $F_{\rm obs} / F_{\rm obs,0} \simeq 0.95$ and $\tau_{\rm fb} / \tau_{\rm dyn} \simeq 3$.  Locating the intersection of the corresponding contours, marked by the (green) open circle in the left panel of Fig.~\ref{fig:darkening}, we identify $q_{\rm min} \simeq 0.5$ and $\sigma_{\rm min} \simeq 3.0$, assuming limb darkening.  Both values correspond to lower limits, since our calculations and the diagrams in (\ref{fig:darkening}) refer to the observed brightness in the $z$-direction, in which one would expect the maximum effect at the moment at which the fly-by object is at the pericenter.  Not knowing the orientation of the observer, the maximum darkening effect might, in fact, be larger than that observed on Earth, meaning that we should find the intersection of  $\tau_{\rm fb} / \tau_{\rm dyn} \simeq 3$ with a value of $F_{\rm obs} / F_{\rm obs,0}$ that is smaller than 0.95, and hence resulting in larger values of $q$ and $\sigma$.

Adopting $M_p = 19 \, M_\odot$ and $R_0 \simeq 750 \, R_\odot$ for Betelgeuse we have $\tau_{\rm dyn} \simeq 0.24$ yrs.  The Great Dimming event lasted about six months, so that $\tau_{\rm fb} / \tau_{\rm dyn} \simeq 2$. During Betelgeuse's Great Dimming, its visual brightness was reduced by approximately one apparent magnitude, or about 60 \%, meaning that $F_{\rm obs}/F_{\rm obs,0} \simeq 0.4$.  Inspecting Fig.~\ref{fig:darkening} we see that it would be very difficult to explain such a large reduction in brightness by means of gravity darkening alone, even under for the most favorable alignment of the fly-by object and observer.  Our estimates suggest that gravity darkening could have led to a reduction in observed brightness of 20 or 30 \%, corresponding to a modest increase in apparent magnitude of about 0.3.  Even such a reduction in brightness would have required $q > 1$, however, meaning that the companion would have had a mass greater than that of Betelgeuse.  While no stellar object was observed close to Betelgeuse during the great dimming, this does not rule out a black hole.

The tidal encounter with such a massive object would also have led to an appreciable exchange of momentum, of course, both a transient exchange during the fly-by, and possibly a permanent exchange.  We may estimate the former by assuming that the relative speed of Betelgeuse and the companion at the pericenter separation, $\varv_s$, is much larger than that at large separation, $\varv_\star$, as we did in Section \ref{sec:timescales}, so that
\begin{equation}
    \varv_s \simeq \left( \frac{2 G M_p}{s} \right)^{1/2} (1 + q)^{1/2}
\end{equation}
(see Eq.~\ref{v_s}).  Betelgeuse's speed at pericenter is then given by
\begin{align}
    \varv_p & = \varv_s \frac{M_c}{M} 
    = \left( \frac{2 G M_p}{s} \right)^{1/2} q \, (1 + q)^{-1/2} \nonumber \\
    & = \left( \frac{2 G M_p}{R_0} \right)^{1/2} \left( \frac{R_0}{s} \right)^{1/2} q \, (1 + q)^{-1/2} = \varv_{\rm esc} \sigma^{-1/2} q \, (1 + q)^{-1/2}, 
\end{align}
where we have introduced Betelgeuse's escape speed $\varv_{\rm esc} \equiv (2 G M_p / R_0)^{1/2} \simeq 100$ km/s. In fact, \cite{Dupetal20} report deviations from Betelguese's average radial velocity during its great dimming on the order of a few km/s.  Adopting such a value for $\varv_p$ we may estimate
\begin{equation}
    \sigma = \frac{s}{R_0} \simeq \left( \frac{\varv_{\rm esc}}{\varv_p} \right)^2 \frac{q^2}{1 + q} \simeq 10^3 \, \frac{q^2}{1 + q}.
\end{equation}
While the changes in radial velocity might indeed point to an interaction with an unseen companion, the above estimate rules out almost all of the scenarios considered in Fig.~\ref{fig:darkening}, and hence suggests that an associated gravity dimming would have been too weak to explain the observed dimming of Betelgeuse.

We caution again that several of our assumptions break down as $\sigma \rightarrow 1$, so that a more careful analysis, e.g., a radiative hydrodynamical simulation, which is well beyond the scope of our simple treatment here, would be necessary to obtain more accurate predictions.

We reiterate that there are other compelling reasons to believe that a gravity darkening effect like the one discussed here was not responsible for the Great Dimming of Betelgeuse -- e.g.~the spectral dependence of the observed darkening.  One of the possible explanations that accommodates these observations is the presence of a large cloud that obscures Betelgeuse from our perspective on the Earth \citep{LevM20,Dupetal20,Monetal21}. 
Such a cloud could have been expelled by Betelgeuse by means of large convective outflows -- but, in fact, might also have been triggered by tidal overflow in a close encounter with an unseen companion.

Finally we note that, while a tidal encounter is unlikely to be responsible for the observed transient brightness change in Betelgeuse, it can trigger such a change in other plausible astrophysical encounters.  For example, if normal stars fly-by supermassive black holes in galaxy cores, the tidal field of the black hole will induce a transient tidal bulge and brightness change in the star.  Whether or not the brightness variation is observable for nearby events (e.g.~Srg A$^*$ stellar fly-bys) is not clear and will depend on the system parameters.  Provided the encounter is not too close to trigger tidal disruption, which occurs less frequently, the analysis provided here may be applicable to such an event.  Similar arguments apply to close binary stars, of course, and light curves from several such systems have indeed been analyzed taking into account the effects of gravity darkening (see, e.g., \cite{RafT80,DjuRRGEP03,DjuRRGEP06} and references therein).  Eta Carinae may also be of interest in this context.  It is known to be in an eccentric binary with a period of 5.5 years that shows spectroscopic and light curve variations correlated with periastron passages (see, e.g., \cite{Dam96,Mehetal15}), so that some of these variations may be caused by gravity-darkening effects.  However, the conventional view is that interacting winds at periastron provide the main explanation for the periodic changes (e.g.~\cite{Kas17} and references
therein).  Further investigation may be warranted to reach and/or refine the correct explanation.

\section*{Acknowledgments}

This work was supported in parts by National Science Foundation (NSF) grant PHY-2010394 to Bowdoin College, as well as NSF grant PHY-2006066 and NASA grant 80NSSC17K0070 to the University of Illinois at Urbana-Champaign.

\section*{Data Availability}

There are no new data associated with this article.

\bibliographystyle{mnras}
% \bibliography{references}

\label{lastpage}

\end{document}